
\input harvmac
\def\inbar{\,\vrule height1.5ex width.4pt depth0pt}
\def\IB{\relax{\rm I\kern-.18em B}}
\def\IC{\relax\hbox{$\inbar\kern-.3em{\rm C}$}}
\def\ID{\relax{\rm I\kern-.18em D}}
\def\IE{\relax{\rm I\kern-.18em E}}
\def\IF{\relax{\rm I\kern-.18em F}}
\def\IG{\relax\hbox{$\inbar\kern-.3em{\rm G}$}}
\def\IH{\relax{\rm I\kern-.18em H}}
\def\II{\relax{\rm I\kern-.18em I}}
\def\IK{\relax{\rm I\kern-.18em K}}
\def\IL{\relax{\rm I\kern-.18em L}}
\def\IM{\relax{\rm I\kern-.18em M}}
\def\IN{\relax{\rm I\kern-.18em N}}
\def\IO{\relax\hbox{$\inbar\kern-.3em{\rm O}$}}
\def\IP{\relax{\rm I\kern-.18em P}}
\def\IQ{\relax\hbox{$\inbar\kern-.3em{\rm Q}$}}
\def\IR{\relax{\rm I\kern-.18em R}}
\font\cmss=cmss10 \font\cmsss=cmss10 at 7pt
\def\IZ{\relax\ifmmode\mathchoice
{\hbox{\cmss Z\kern-.4em Z}}{\hbox{\cmss Z\kern-.4em Z}}
{\lower.9pt\hbox{\cmsss Z\kern-.4em Z}}
{\lower1.2pt\hbox{\cmsss Z\kern-.4em Z}}\else{\cmss Z\kern-.4em Z}\fi}
\def\IGa{\relax\hbox{${\rm I}\kern-.18em\Gamma$}}
\def\IPi{\relax\hbox{${\rm I}\kern-.18em\Pi$}}
\def\ITh{\relax\hbox{$\inbar\kern-.3em\Theta$}}
\def\IOm{\relax\hbox{$\inbar\kern-3.00pt\Omega$}}

\def\NP{Nucl. Phys. }
\def\PL{Phys. Lett. }
\def\PRL{Phys. Rev. Lett. }
\def\PRV{Phys. Rev. }

\def\PR{Phys. Rep. }

\def \CO{{\cal O}}

\def \CH{{\cal H}}

\def\semidirect{\mathbin{\hbox{\hskip2pt\vrule height 4.1pt depth -.3pt
width .25pt \hskip-2pt$\times$}}}

\Title{RU-93-39, hep-ph/9308363}{\vbox{\centerline{(S)quark Masses}
\vskip2pt\centerline{and Non-Abelian Horizontal Symmetries}}}
\bigskip
\centerline{\it Philippe Pouliot and Nathan Seiberg}
\smallskip
\centerline{Department of Physics and Astronomy}
\centerline{Rutgers University}
\centerline{Piscataway, NJ 08855-0849}
\noindent
\bigskip
\baselineskip 18pt
\noindent
We present a model of quark and squark masses which is based on a
non-Abelian horizontal symmetry.  It leads to order of magnitude
relations between quark mass ratios and mixing angles and to the
successful exact relation $\sin \theta_C=\sqrt {m_d\over m_s}$ to better
than $20\%$ accuracy.  The non-Abelian symmetry also ensures the
necessary squark degeneracy to suppress FCNC mediated by loops with
squarks and gluinos, in the neutral meson systems.

\Date{8/93}
One of the most puzzling open questions in particle physics is the
origin of the small numbers in the Standard Model Lagrangian.  These
numbers include the mixing angles and the ratios of quark masses.
Typically in physics, small numbers are associated with an explicitly
broken symmetry which becomes exact when these numbers are set
to zero~\ref\thooft{G. 'tHooft, Lecture at the Cargese Summer Institute,
(1979).}.
Several authors have suggested to use a horizontal symmetry $\CH$,
which acts on the quarks and the leptons, as the underlying symmetry
controlling the small numbers (for a recent discussion, see e.g.
references~\ref\hall{A. Antaramian,
L.J. Hall and A. Rasin, \PRL {\bf 69} (1992)
1871; L.J. Hall and S. Weinberg,  \PRV {\bf D48} (1993) R979.}).
Such an explicitly broken horizontal symmetry can arise
naturally~\ref\lnsn{M. Leurer, Y. Nir and N. Seiberg, to appear.}
when an exact symmetry $\CH$ is spontaneously broken by an expectation
value of some scalar field $\vev{S}$.  The small numbers in the
Lagrangian then appear as powers of the ratio $\lambda = {\vev S \over
M}$, where $M$ is the scale of higher energy physics which communicates
the information about $\CH$ breaking to the quarks.  Explicit mechanisms
based on massive fermions with mass $M$ were suggested in
references~\ref\frni{C.D. Froggatt and H.B. Nielsen, \NP {\bf B147}
(1979) 277.}~\ref\dimo{S. Dimopoulos, \PL {\bf 129B} (1983) 417;
J. Bagger, S. Dimopoulos, E. Masso and M. Reno, \NP {\bf B258} (1985)
565;
J. Bagger, S. Dimopoulos, H. Georgi and S.~Raby, In: {\it Proc. Fifth
Workshop on Grand Unification.} Eds. Kang, K., Fried, H. and Frampton,
P., Singapore, World Scientific (1984).}~\ref\lns{M. Leurer,
Y. Nir and N. Seiberg, \NP {\bf 398B} (1993) 319.}.
For simplicity, we rescale $S$ and set~$M=1$.

The simplest $\CH$ is Abelian and it easily leads to order of magnitude
relations between quark mass ratios and angles~\lns~\lnsn.  These
explain the Wolfenstein parametrization of the CKM matrix
\eqn\wolf{V_{\rm CKM}=
\pmatrix{1-{\lambda^2\over 2}& \lambda & \lambda^3A(\rho+i\eta)\cr
-\lambda & 1- {\lambda^2\over 2} & \lambda^2A \cr
\lambda^3A(1-\rho+i\eta) & -\lambda^2A & 1 \cr} }
with $A \sim 1$, $\lambda \sim 0.2$ and the mass ratios and phases
\eqn\massra{\eqalign{ m_u/m_c\sim & \lambda^3 - \lambda^4 \qquad
\qquad m_c/m_t\sim \lambda^3 - \lambda^4 \cr
m_d/m_s \sim & \lambda^2 \qquad
\qquad\qquad m_s/m_b \sim \lambda^2 \cr
m_b/m_t \sim & \lambda^2 -\lambda^3 (\lambda^3)
\qquad \rho, \eta \sim 1 -\lambda \quad . \cr}}
These estimates are valid in the TeV range (when the expressions at high
energy, such as $10^{15}\, {\rm GeV}$, are different, we have indicated
them
in parentheses).  Here and throughout this work we take $\tan \beta =
{\vev{H_u} \over \vev{H_d}} \sim 1$.  Note that order of magnitude
estimates as those in equation~\massra\ are ambiguous and sensitive to
the exact value of $\lambda$.

A better theory should be more predictive.  One would like to find some
exact or at least approximate relations and not only order of magnitude
ones.  Such relations can arise {}from a more detailed knowledge of the
high energy theory.  For example in grand unified theories there are
some exact relations among quark mass ratios and mixing
angles~\ref\exactgrand{H. Georgi and C. Jarlskog, \PL { \bf 89B} (1979)
297;
J.A. Harvey, P. Ramond and D.B.~Reiss, \PL {\bf 92B} (1980) 309; \NP
{\bf B199} (1982) 223;
S. Dimopoulos, L. J. Hall and S. Raby, \PRL {\bf 68} (1992) 1984.}.
The approach we will take here is based on a non-Abelian $\CH$.  The
group theory of $\CH$ can fix some of the couplings of order one as
Clebsch-Gordan coefficients and lead to exact relations.  A successful
exact relation which we will try to explain is
\eqn\famous{\sin \theta_C=\sqrt {m_d\over m_s}}
which is satisfied at about $5-10\%$, namely to $\CO (\lambda -
\lambda^2)$.  Recent CLEO results for
$|V_{ub}/V_{cb}|$~\ref\bsg{E. Thorndike, CLEO Collaboration, talk
given at the meeting of the American Physical Society, Washington D.C.
(1993).}
may favor $\rho$, $\eta\sim\lambda$, which would lead to another
close-to-exact relation
\eqn\anotherrel{\vert V_{td}\vert =\vert V_{cb}V_{us}
\vert  [1+\CO(\lambda)]~.}

A related issue is the squark spectrum.  It is constrained by
FCNC~\ref\squade{R. Barbieri and R. Gatto, \PL {\bf 110B} (1982) 211;
J. Ellis and D.V. Nanopoulos, \PL {\bf 110B} (1982) 44;
H.P. Nilles, \PR {\bf 110} (1984) 1;
F. Gabbiani and A.~Masiero, \NP {\bf B322} (1989) 235.}.
Usually these constraints are satisfied by postulating that all squarks
are degenerate.  In standard supergravity or string theory this
degeneracy is not natural.  However, under special circumstances in
string theory the squarks might be naturally
degenerate~\ref\vadim{V.S. Kaplunovsky and J. Louis, \PL {\bf B306}
(1993) 269;
R. Barbieri, J.~Louis and M. Moretti, \PL {\bf B312} (1993) 451.}.
Also, if supersymmetry breaking is fed to the squarks by QCD
interactions, the squarks are degenerate~\ref\oldnewdine{M. Dine,
A. Kagan and S. Samuel, \PL {\bf B243} (1990)
250; M. Dine and A.~Nelson, \PRV {\bf D48} (1993) 1277.}.
In reference~\ref\nsei{Y. Nir and N. Seiberg, \PL {\bf B309} (1993) 337.}
it was suggested that FCNC constraints can be satisfied with
non-degenerate squarks by arranging that the quark and squark mass
matrices be approximately diagonal in the same basis.  This can be
achieved with an Abelian $\CH$.  An alternative, which we will follow
here is to use a non-Abelian $\CH$ to guarantee the degeneracy of some
of the squarks~\ref\dkl{M. Dine,
A. Kagan and R. Leigh, SCIPP 93/04/SLAC-PUB-6147,
Mar. 1993.},
along with alignment.
\bigskip
The goal of this work is to use a non-Abelian horizontal symmetry $\CH$
which

\noindent
1. explains the order of magnitude relations \wolf\ and \massra\ as in
references~\lns\lnsn,

\noindent
2. leads to the approximate relations \famous\ and \anotherrel,

\noindent
3. guarantees the degeneracy of the squarks $\tilde d$ and $\tilde s$
and the degeneracy of $\tilde {\bar d}$ and $\tilde {\bar s}$ to be
compatible with the data on $K-\bar K$ mixing as in reference~\dkl\ and

\noindent
4. aligns the quark and squark mass matrices as in
references~\nsei~\lnsn\ to be consistent with all other FCNC data.

Below we will present an example satisfying these requirements.  Our
example is not simple and is certainly not unique.  It should be thought
of as an existence proof that such a program is possible.

A few general arguments guide us in the search for a model.

We will use the following notation for the superfields.  $Q_i$ denote
the left-handed doublets and $\bar d_i$ ($\bar u_i$) denote the
left-handed anti-down (anti-up) $SU(2)$ singlets.  The mass matrices are
$M^d$ and $M^u$.

Since the top mass is of the same order of magnitude as the Higgs VEVs,
$M^u_{3,3}$ should not be suppressed by any power of the small parameter
$\lambda$.  Therefore, using the baryon number symmetry the fields $Q_3$
and $\bar u_3$ can be taken to transform trivially under $\CH$.

To obtain equation~\famous\ as a prediction we should use $\CH$ to
ensure the equality of $M^d_{1,2}$ and $M^d_{2,1}$, and make $M^d_{1,1}
\ll (M^d_{1,2})^2/M^d_{2,2}$.  Also, we should make sure that the
contribution to $V_{us}$ {}from the up sector is sufficiently small.
This can be achieved when $Q=(Q_1, Q_2)$ transforms as a doublet of $\CH$
and either $\bar d=(\bar d_1, \bar d_2)$ is a doublet and $\bar d_3$ a
singlet or $\bar d=(\bar d_1,\bar d_2,\bar d_3)$ is a triplet.  We will
focus on the first of these possibilities.  This choice of
representations also guarantees the squark degeneracy mentioned above.

The $\bar u$ quarks could all be (perhaps non-trivial) singlets or $\bar
u= (\bar u_1,\bar u_2)$ could be a doublet of $\CH$.  We will focus on
the latter possibility, which also leads to the degeneracy of the
squarks $\tilde { \bar u}$ and $\tilde {\bar c}$.

As explained in reference~\lns, at least one of the standard model
singlet fields $S$ has to be in a multi-dimensional irreducible
representation of $\CH$; we take it to be a doublet.  We also found it
helpful to add an extra Standard Model singlet field $T$ to explain
$M^d_{3,3} \ll M^u_{3,3}$ when $\tan \beta \sim 1$.
\bigskip
We now consider our explicit example.  We start with the group $U(1)
\times O(2)$ and label the representations in terms of the charges under
the two factors\foot{$O(2)\cong Z_2\semidirect U(1) $ has
two-dimensional representations labeled by a positive integer charge
under the $U(1)$ subgroup and two different one-dimensional
representations with vanishing charge and different $Z_2$
assignments.}.  We assign the quarks and scalars to
\eqn\firstmodel{\matrix{Q & \bar d & \bar d_3 & \bar u & S & T \cr
(8,2) & (16,2) & (12,0) & (11,1) & (-5,1) & (-4,0) \cr}}
where  $\bar d_3$ and $T$ transform trivially under $O(2)$.  Since both
charges have the same parity, only $G=(U(1) \times O(2))/Z_2$ is
represented\foot{Note that $G\cong Z_2\semidirect(U(1)\times U(1))$
which suggests a connection to string inspired models.  Non-Abelian
discrete groups of the form $S_m \semidirect (Z_n \times Z_n\cdots
Z_n)$ with $m$ factors of $Z_n$ are  common in Calabi-Yau
compactifications.  For $m=2$ this is a discrete subgroup of our $G$.}.
(The $U(1)$ charges in equation \firstmodel\ are smaller and look more
natural, if the $U(1)$ group is replaced by its $Z_9$ subgroup.  This
does not affect our results.)

The Yukawa terms in the Lagrangian lead to the mass matrices
\eqn\downmassmatrices{M^d= y_d \vev{H_d}
\pmatrix{TS_2^4 & T^6 \!  + \! \eta TS_1^2S_2^2 & \gamma S_1S_2^3\cr
T^6\! +\! \eta TS_1^2S_2^2 & TS_1^4 & \gamma S_1^3S_2\cr
0 &0 & T^3 \cr} }
\eqn\upmassmatrices{M^u= y_u \vev{H _u}
\pmatrix{\alpha TS_2^3 & \beta TS_1S_2^2  & 0 \cr
\beta TS_1^2S_2& \alpha TS_1^3 & 0 \cr
0 &0 & 1 \cr} }
here we have used a basis for the two-dimensional representations in
which the $U(1)$ subgroup of $O(2)$ is diagonal.  In writing these
equations two of the Yukawa coefficients have been absorbed into the
definitions of the fields $S$ and $T$.  (We remind the reader that we
have also divided the fields $S$ and $T$ by the scale~$M$ at which
symmetry breaking is communicated to the quarks.)

A scalar potential which we do not discuss leads to expectation values
to the two scalars in $S$ and to the scalar $T$.  The three expectation
values can be different and lead to three stages of symmetry breaking.
Instead, we assume that first $S_1$ and $T$ acquire expectation values
of order $\lambda \sim 0.2$ and break $G$ to $Z_2$ which is broken at a
lower scale by $\vev{S_2} \sim \lambda^2 \sim 0.04$.  (In the version of
the model based on $Z_9 \subset U(1)$ this pattern of symmetry breaking
is natural (i.e. $\vev{S_2} \ll \vev{S_1}$, $\vev{T}$ is protected by a
symmetry) only when another $Z_4$ group under which $S$ and $\bar u$
have charge one is added.)

To leading order in $\lambda$, the mass matrices of equations
\downmassmatrices\ and \upmassmatrices\ lead to the mass
ratios\foot{Off-diagonal wave function renormalization which depends on
$\lambda$ should also be taken into account
\ref\ds{A. Dabholkar and N. Seiberg, unpublished.}
\lnsn.  However, because of the quantum numbers of the fields in our
model this effect is negligible.}
\eqn\relations{\eqalign{
{m_s\over m_b}\cong &\> {S_1^4 \over T^2} \sim \lambda^2
\qquad {m_c \over m_t}\cong \alpha\, TS_1^3 \sim \lambda^4 \cr
{m_d\over m_s} \cong & \> {T^{10} \over S_1^8} \sim \lambda^2
\qquad {m_u\over m_c}\cong \left(1-{\beta^2\over
\alpha^2}\right) {S_2^3\over  S_1^3} \sim \lambda^3 \cr
{m_b\over m_t} \cong & \> {y_d \, T^3 \over y_u \tan \beta} \sim
\lambda^3\cr}}
and the mixing angles
\eqn\mixangle{\eqalign{
\vert V_{us}\vert = & \> {T^5 \over S_1^4} + \CO\left( {S_2^2\over
S_1^2}\right) = \sqrt{m_d\over m_s}\,  [1+ \CO(\lambda)] \sim \lambda
\cr
\vert V_{cb} \vert \cong &\>  {\gamma S_1^3 S_2 \over T^3}\sim \lambda^2
\cr
\vert V_{ub}\vert \cong &\>  {\gamma S_1 S_2^3 \over T^3} \left(1-{\beta
\over \alpha}\right) \sim \lambda^4  \cr
\vert V_{td}\vert = & \> \vert V_{us}V_{cb} \vert
[1+ \CO(\lambda)] \sim \lambda^3
\quad. \cr}}

The (hermitian) squark mass matrices are $6 \times 6$ matrices.  The
off-diagonal $3\times 3$ blocks are proportional to $\tilde m \vev
{H_{u,d}}$ where $\tilde m \sim 1\, {\rm TeV}$ is a typical squark mass
and are similar to the quark mass matrices (in order of magnitude but
not necessarily the same coefficients).  If $\tilde m^2$ is much larger
than the entries in the off-diagonal blocks, we can focus on the
diagonal $3 \times 3$ blocks.  The leading order contributions to the
mass square matrices are
\eqn\tildeqm{ \tilde M^2_Q = \tilde m^2  \tilde \delta^Q =  \tilde m^2
\pmatrix{a_Q  & c_Q S_2^2S_1^{*2} & d_Q S_2S_1^*T^2 \cr
& a_Q  & d_Q S_1S_2^*T^2 \cr
& & b_Q \cr}
\sim \tilde m^2 \pmatrix{1 & \lambda^6 & \lambda^5 \cr
\lambda^6 & 1 & \lambda^5 \cr
\lambda^5 & \lambda^5 & 1 \cr}
}
\eqn\tildedm{\tilde M^2_{\bar d} = \tilde m^2 \tilde
\delta^{\bar d} =
\tilde m^2 \pmatrix{a_d& c_d S_2^2S_1^{*2} & d_d S_2S_1^*T \cr
& a_d & d_d S_1S_2^*T\cr
& & b_d \cr}
\sim \tilde m^2 \pmatrix{1 & \lambda^6 & \lambda^4\cr
\lambda^6 & 1 & \lambda^4 \cr
\lambda^4 & \lambda^4& 1 \cr}
}
\eqn\tildeum{\tilde M^2_{\bar u} =\tilde m^2 \tilde
\delta^{\bar u} = \tilde m^2
\pmatrix{a_u & c_u S_2S_1^* & d_u S_1^*T^4\! + \! d^\prime_u S_1S_2^2T^*
\cr
& a_u & d_u S_2^*T^4\!  + \! d^\prime_u S_1^2S_2T^*\cr
& & b_u\cr}
\sim \tilde m^2 \pmatrix{1 & \lambda^3 & \lambda^5\cr
\lambda^3& 1 & \lambda^5 \cr
\lambda^5 & \lambda^5  & 1 \cr}
}
where the coefficients $a_i$, $b_i$, $c_i$, $d_i$, $d^\prime_u$ are
constants of order one.  Note that the first two diagonal entries are
equal up to corrections proportional to $(S_1S_1^*- S_2S_2^*)$ which are
of order $\lambda^2$.  The third diagonal entry can be different.

The FCNC constraints on the squark mass matrices are best presented in
the basis related by supersymmetry to the basis where the quark mass
matrices are diagonal.  We transform the matrices $\tilde \delta$ in
equations \tildeqm, \tildedm\ and \tildeum\ to this basis and find four
new matrices $\delta^u$, $\delta^d$, $\delta^{\bar u}$ and $\delta^{\bar
d}$.  The values of the experimental upper bounds and the order of
magnitude predictions in our model are
\eqn\dfcnct{\matrix{
(\delta^d_{ds} \delta^{\bar d}_{ds})^{1/2}  & \delta^d_{ds} &
\delta^{\bar d}_{ds} & (\delta^d_{db} \delta^{\bar d}_{db})^{1/2}
&\delta^d_{db} &\delta^{\bar d}_{db}& (\delta^u_{uc} \delta^{\bar
u}_{uc})^{1/2} & \delta^u_{uc} & \delta^{\bar u}_{uc} \cr
0.006 & 0.05 & 0.05 & 0.04 & 0.1 & 0.1 & 0.04 & 0.1 & 0.1 \cr
\lambda^3 & \lambda^3 & \lambda^3 &\lambda^{3.5} & \lambda^3 &
\lambda^4 & \lambda^{3.5} & \lambda^4 & \lambda^3 \cr} \quad .}
The experimental upper bounds suffer {}from multiplicative ambiguities
of order 3 -- 4.  Similarly, there is also a factor of order one
ambiguity in these predictions of our model which arises {}from the
coefficients we do not know.

The bound on $K-\bar K$ mixing is satisfied because the squarks in the
first two generations have sufficiently degenerate masses.  $B-\bar B$
and $D-\bar D$ mixing bounds are satisfied because of alignment~\nsei:
in the basis where the quark mass matrices are diagonal, the squark mass
matrices are also approximately diagonal.

CP violation in the K system constrains ${\rm Im}~\delta^d_{ds}$ and
${\rm Im}~\delta^{\bar d}_{ds}$ to be about an order of magnitude below
the values quoted in equation \dfcnct.  If all phases in the various
mass matrices are of order one, the leading contribution to these
imaginary parts is of order $\lambda^3$ which is too large.  Therefore,
to be consistent with the experimental value of $\epsilon_K$ some of the
phases should be small.  This can arise only from a better understanding
of the origin of CP violation.

So far we have not discussed all the terms in the Lagrangian.  Without
the ``$\mu$ term'' $\mu H_u H_d$, the theory has another $U(1)_{PQ}$
symmetry.  The horizontal symmetry can be taken as any subgroup $\CH
\subset G \times U(1)_{PQ}$.  One should only make sure that the extra
terms which are allowed by $\CH$ do not ruin our results.  The group
$\CH$ can be continuous or discrete.  If it has discrete factors, one
needs to make sure that there is no cosmological problem with domain
walls when $\CH$ is broken.

If squark masses are ever measured, a number of patterns can be found:

\noindent
1. All twelve squarks are degenerate.  The explanation of this fact is
probably unrelated to the origin of quark masses.

\noindent
2. The squark pairs $(\tilde d, \tilde s)$, $(\tilde {\bar d}, \tilde
{\bar s})$, $(\tilde u, \tilde c)$, $(\tilde {\bar u}, \tilde {\bar c})$
are degenerate.  The different pairs are not necessarily degenerate and
neither are the other squarks.  Such a situation arises in models with a
horizontal symmetry where $\bar u$ is in a doublet of $\CH$ as in
the model of reference~\dkl\ and in our model.

\noindent
3.  The degeneracy is as in the previous case except that the pair
$(\tilde {\bar u}, \tilde {\bar c})$ is not degenerate.  This can be a
signal that $\bar u$ is in a reducible representation of $\CH$.  For
example, we can replace the doublet $\bar u$ in our model with two
singlets $\bar u_1$ and $\bar u_2$ transforming under $G$ like $(12,0)$
and $(10,0)$ respectively and find a model with a massless up quark.
In this model, $D-\bar D$ mixing is suppressed by one more power of
$\lambda$ relative to the aligned models of reference \nsei.  However,
it is still much larger than predicted by the Standard Model or by
supersymmetric models where all squarks are degenerate.  This model is
less constrained in the quark sector, allowing for smaller $m_u/m_c$ and
larger $V_{ub}$ than the previous model.

\noindent
4. No two squarks are degenerate as in the aligned models of
references~\nsei\lnsn.

To summarize, our model is natural.  We include in the Lagrangian all
terms consistent with the symmetry with coefficients of order one. $\CH$
is broken in two stages.  At the first step all scalars invariant under
the unbroken symmetry acquire VEVs and the other scalars are protected.

The only small parameters are in the VEVs of the singlets $S$ and $T$
which break~$\CH$.  By raising them to various powers which are
determined by $\CH$, we generate all the small parameters of the squark
sector.  We find several order of magnitude relations as well as the
approximate relation $\sin \theta_C=\sqrt{m_d\over m_s}$ at the $20\%$
level (order $\lambda$).  With an appropriate choice of Yukawa couplings
of order one, our results are consistent with all known phenomenology of
masses and mixing angles.  Also, as in reference~\dkl, we explain some
of the degeneracy among the squarks thus ensuring the suppression of
FCNC.

\centerline{\bf Acknowledgments}
\nobreak
It is a pleasure to thank T.~Banks, A.~Dabholkar, K.~Davis, M.~Dine,
K.~Intriligator, D.~Kaplan and A.~Nelson and especially M.~Leurer
and Y.~Nir for several useful discussions.  We also thank Y.~Nir for
comments on the manuscript.  This work was supported in part by
DOE grant DE-FG05-90ER40559 and a Canadian 1967 Science fellowship.
\listrefs
\bye